\documentclass[12pt]{article}   
 \usepackage{psfig,times,fancyheadings}
        
\begin{document}
\thispagestyle{empty} 

% \lhead[\fancyplain{}{\sl Price waves}]{\fancyplain{}{\sl The    
% Newsletter  of Econophysics}}
% \rhead[\fancyplain{}{\sl The Newsletter of Econophysics}]{\fancyplain{}{\sl   
% April 1997 Number 1}}

 \lhead[\fancyplain{}{\sl }]{\fancyplain{}{\sl }}
 \rhead[\fancyplain{}{\sl }]{\fancyplain{}{\sl }}

%%%%%%%% Pour changer les valeurs par defaut pour taille figure,
%%%%%%%% sinon au-dela d'une hauteur de 134 mm = 70% on est rejete a la fin
 \renewcommand{\topfraction}{.99}      
 \renewcommand{\bottomfraction}{.99} 
 \renewcommand{\textfraction}{.0}

%%%%% Definitions

\newcommand{\nc}{\newcommand}

\nc{\qI}[1]{\section{{#1}}}
\nc{\qA}[1]{\subsection{{#1}}}
\nc{\qun}[1]{\subsubsection{{#1}}}
\nc{\qa}[1]{\paragraph{{#1}}}

            % Enumerations
\def\qbu{\hfill \par \hskip 6mm $ \bullet $ \hskip 2mm}
\def\qee#1{\hfill \par \hskip 6mm #1 \hskip 2 mm}

\nc{\qfoot}[1]{\footnote{{#1}}}
\def\qL{\hfill \break}
\def\qpar{\vskip 2mm plus 0.2mm minus 0.2mm}
\def\tvi{\vrule height 12pt depth 5pt width 0pt}

\def\qparr{ \vskip 1.0mm plus 0.2mm minus 0.2mm \hangindent=10mm
\hangafter=1}

                % Decale UN paragraphe
                % Attention! La double accolade est vitale, sinon tout le
                % est decale (cf TEX p.199)
                % On peut aller a la ligne avec \qL=\hfill \break
                % Par contre ne supporte pas les lignes blanches
\def\qdec#1{\par {\leftskip=2cm {#1} \par}}

   %% Defs specifiques
\def\qdpt{\partial_t}
\def\qdpx{\partial_x}
\def\qddpt{\partial^{2}_{t^2}}
\def\qddpx{\partial^{2}_{x^2}}
\def\qn#1{\eqno \hbox{(#1)}}
\def\qds{\displaystyle}
\def\qw{\widetilde}
\def\qmax{\mathop{\rm Max}}   % Petit livre Tex (p.167)
\def\qmin{\mathop{\rm Min}}   % Petit livre Tex (p.167)

%%%%% End of definitions

\null

% Interligne plus large pour faciliter la relecture (corrections)
%\baselineskip=3mm

\vskip 0.5cm
 {\bf \sl \large To appear in: International Journal of Modern Physics C} 
\vskip 1cm

\centerline{\bf \LARGE To sell or not to sell?}
\vskip 0.5cm
 \centerline{\bf \LARGE Behavior of shareholders during price collapses}

\vskip 1cm
\centerline{\bf Bertrand M. Roehner $ ^1 $ }
\centerline{\bf L.P.T.H.E. \quad University Paris 7 }

\vskip 2cm

{\bf Abstract}\quad  It is a common belief that the behavior of
shareholders depends upon the direction of price fluctuations: if
prices increase they buy, if prices decrease they sell. That belief,
however, is more based on ``common sense'' than on facts. In this
paper we present evidence for a specific class of shareholders
which shows that the actual behavior of shareholders can be
markedly different. For instance, they may continue to 
buy despite a prolonged fall in prices or they may sell
even though prices climb. A closer analysis shows that 
a substantial proportion of investors
are more influenced by the ``general social climate'' than by 
actual price changes. The percentage of speculative investors 
who optimize their portfolio on a monthly basis can be estimated
and turns out to be about 5 to 10 percent.
The results presented in this paper can be of usefulness in order
to test the assumptions or the results of market simulations
and models.

\vskip 1cm
 \centerline{\bf  27 January  2001 }
%\centerline{\bf  8 October  2000 \quad First draft}

\vskip 1cm
% {\bf PACS.}\ 
% 64.60 Equilibrium properties near critical points - 
% 87.23Ge Dynamics of social systems 

Keywords: Stock prices, supply and demand, equity funds, 
          inflow-outflow

\vskip 2cm
1: Postal address: LPTHE, University Paris 7, 2 place Jussieu, 
75005 Paris, France.
\qL
\phantom{1: }E-mail: ROEHNER@LPTHE.JUSSIEU.FR
\qL
\phantom{1: }FAX: 33 1 44 27 79 90

\vfill \eject

\qI{Introduction}

%%%%%%%%%%%%%%%

   \count101=0
   \ifnum\count101=1
\qdec{``[In April 2000] the flood of money into equities [i.e. stocks]
has slowed to a trickle. Compared with March new money invested in
equity mutual funds fell by half last month to 6 billion dollars
a week. The amazing thing is that it continues at all.''}
We placed this excerpt from the Economist (27 May 2000) at the 
beginning of this paper in order to show how far of the mark
commonly held opinions can be, regarding the way investors
react to a market crash. For one thing, the inflow of new money
into equity funds by no means ``slowed to a trickle''; it is true
that the figure for April 2000 was lower than the one for March,
but it was still larger than the figure for April 1999 (see table 1a).
Although the inflow again slowed in May it was still higher than
in the same month of 1999. Moreover the astonishment expressed by
the writer in his (her) last observation shows that the common
view is that, in the age of on line trading, investors react 
to market news within days or even within minutes. As will be
shown in this paper they rather react with a delay which can
be expressed in years. 
\qL

  \fi
%%%%%%%%%%%%%%%%

\qdec{``[Between March and May 2000] the hugely popular OTC and Emerging
fund-management firm has lost 55 percent of its value; and yet new
money continues to be invested in it, albeit at the rate of
50 million dollars a week, instead of some 200 million dollars.''}
\qL
We placed this excerpt from the Economist (27 May 2000) at the 
beginning of this paper in order to show on a specific example
that even in the age of Internet communication and on line trading 
not all 
shareholders react to adverse news within days. One of the main
objectives of this paper is precisely to estimate the proportion of
the quickly reacting investors.
\qL
In order to fulfill that program we must focus on a specific class of
shareholders. Indeed, to our knowledge, 
the only published figures
regarding flows of money into or out from
stock markets are those for 
investors who own stocks through mutual funds.
Subsequently such shareholders will be
referred to as mutual fund shareholders or more shortly as
mf shareholders. The fact that one can
know the net flow of money in the weeks, months or years
following a stockprice collapse
is of central importance for the present study. 
Subsequently in-flows will 
be counted positively and out-flows negatively. 
\qpar

To begin with, let us briefly give some general information
about American mutual funds. In 1999 there were about 8,000
funds of which bond funds represented 30 percent and equity
funds (also called stock funds) about 50 percent (Mutual
Fund Factbook 2000). The total assets of equity funds represented
2.4 trillion dollars in 1997 that is to say 25 percent of the
capitalization of the New York Stock Exchange (NYSE) or
20 percent of the combined capitalization of the NYSE and NASDAQ
(Statistical Abstract of the United States 1999). 
These orders of magnitude show that the impact of mf shareholders
is far from being marginal. 
The mf shareholders are either households or institutions; in
1999 the latter (mainly retirement funds) represented one third
of the assets.

\qA{Formulation of the problem}
Table 1a gives monthly data for in-flows of new money into 
equity funds for the years 1997-2000. But how should these data
be interpreted? That is what we want to discuss in the present
paragraph. In order to simplify the discussion we consider 
the two following sorts of investors.
\qbu The class of what we call speculative investors  who
optimize their investments on a daily or weekly basis. For instance,
if the return of stock markets in a given month is negative while
at the same time
it is possible to get an annual 6 percent interest on bond markets
they will make an arbitrage in favor of the later and shift 
a substantial part of their assets to the bond market. 
\qbu The class of what we call long-term investors who make their
decisions on a multi-annual basis. 
\qL
In real life there is probably a whole spectrum of attitudes 
between these borderline cases. It is mainly for the sake of 
simplicity that we restrict ourselves to these two classes.
A convenient way to describe an actual population in that 
simplified framework is to consider that it is a mix of speculative
and long-term investors. One of the key variables in the present
study is the percentage represented by each class and one of 
our main goals is to get reliable estimates for them.
\qpar

Now let us see on the specific example of the monthly data for 2000
how the above framework can be put to use. Two observations
are in order. (i) For almost all months
(with the only exception of August) the yields of bonds (either
short-term or long-term) were higher than stock market returns.
Thus, if all mf shareholders had been speculative investors they 
would have channeled their money into the bond market (or toward
other lucrative markets such as for instance the real estate market
which experienced a boom in 2000). In other words the in-flow of
new money would have stopped or would even have become negative.
In contrast the evidence shows that there was
a total net in-flow of 290 billion dollars for 2000, which 
proves that some mf shareholders are rather long-term investors. 
\qL
(ii) If one computes the regression between stock price changes
($ \Delta p/p $) and net in-flows of money ($ f $) one obtains
(correlation is 0.57):
$$ f = a\Delta p/p + b  \quad a=1.36 \pm 1.35 \quad b=25 \pm 7 \qn{1} $$

which shows that there is a definite relationship between flows and
price changes. Incidentally it can be noted that whereas
the error margin for $ a $ is fairly high in the 
present case 
(this is because the regression is made on only 10 points)
it is somewhat smaller for previous years (see table 1b). 
Equation (1) shows that the population of shareholders does not 
comprise 100 percent long-term investors 
(otherwise $ f $ would not change with $ \Delta p/p $). 
\qpar

More specifically it is possible to derive an estimate
of the proportion $ \alpha $ of speculative investors 
from equation (1). 
Indeed, under the assumption of two classes of investors 
the flow $ f $ would be given by:
$$ f/k = \alpha (\Delta p/p) + (1-\alpha ).1  \qn{2} $$

$ k $ denotes a proportionality constant between $ f $ and the price
changes; if $ \alpha =1 $ equation (2) reduces to: 
$ f/k = \alpha \Delta p/p $ which means that the in-flows of money
would be completely controlled by price changes; on the contrary if
$ \alpha =0 $ they would be independent of price changes. 
\qL
Identifying equations (1) and (2) gives:
$$ { a \over \alpha } = { b \over 1- \alpha } \Longrightarrow
\alpha = { a/b \over 1+ a/b } $$

With the data for 2000 this leads to: 
$ \alpha = 0.054 /(1+0.054) = 5.1 $ percent. 
\qL
Incidentally it can be observed that if $ a/b $ is small with respect
to 1 one has approximately: $ \alpha = a/b $. 
\qL
For other years (table 1b) one obtains estimates for $ \alpha $ which are
in the (5 \%, 10 \%) range. Such an
order of magnitude is consistent with
what is known about the behavior of mf shareholders through surveys
(see below).

\qA{Some methodological points}
In the previous paragraph a number of methodological options have
been selected which deserve a more detailed discussion. 
\qL
(i) In considering portfolio optimization we neglected 
transactions costs. Was that justified?
For stocks, transaction costs have notably declined in the 1990s
(for more details see Roehner 2001, p.115); in the late 1990s they 
were of the order of one cent per share that is to say
about 0.2 percent. For equity funds the decline was much slower
however. In 1990 transaction costs represented
about 1.8 percent and by 1998 they had declined
to 1.3 percent (Reid 2000, p.18). 
Since the return differential between stocks and bonds
was in 2000 at least 6 percent, even such 
high transaction costs cannot deter
speculative investors from moving their money from one market
to another; however it certainly induces much friction and partly 
explains the reluctance of mf shareholders to switch to a different
investment strategy. 
\qL
(ii) In so far as equity funds comprise American as well as foreign
stocks one may wonder if one should not use a world index for stock
prices. In fact, throughout the 1990s foreign stocks represented
only about 10 percent of the total assets of American equity funds:
the figure was 11 percent in 1990 and 13 percent in 1999 (Reid 2000,
p.14). 
\qL
(iii) One may wonder why the stock prices and money flows in table
1a were not corrected for inflation. In fact such a correction 
would be quite negligible: during the 1990s the annual inflation
rate was on average less than 2 percent which means that on a monthly
basis it was less than 0.2 percent. Such a correction
is far smaller than the error margin on the flow data. However,
when subsequently we consider yearly data over the whole decade
we use deflated figures (over the decade the price increase was about
30 percent).
\qL
(iv) A last point concerns the fact that we used $ f $ as our 
dependent variable instead of $ \Delta f/f $  or some other
measure of fluctuations such as the difference between monthly values
and the overall trend (for instance in the form of a moving average).
In fact the regressions were also carried out for $ \Delta f/f $,
with basically similar results. We preferred using
$ f $ because this is already an increment: it represents the
change in the funds assets, $ \Delta f/f $ would be a second-order 
increment. Furthermore to use the difference between monthly 
values and a trend would mean that we renounce to explain the
trend; this would simply disregard the low-frequency response
of long-term investors.

 % TABLE 1a

\begin{table}[htb]
\footnotesize

\centerline{\bf \small
Table 1a \ Net new money invested in equity mutual funds}
\centerline{\bf \small versus stock price changes, monthly data}

\vskip 3mm
\hrule
\vskip 0.5mm
\hrule
\vskip 2mm
$$ \matrix{
\tvi  &\hbox{Jan.}&\hbox{Feb.}&\hbox{Mar.}& 
\hbox{Apr.}&\hbox{May.}&\hbox{Jun.}&\hbox{Jul.}&\hbox{Aug.}&
\hbox{Sep.}&\hbox{Oct.}&\hbox{Nov.}&\hbox{Dec.}\cr
\noalign{\hrule}
\tvi  \hbox{\bf 1997}  & & & & & & & & & & & & \cr
\hbox{NY+NAS (\%)} \hfill & & -0.23&-4.6&4.3&7.0&4.4&7.6& -3.9
&5.8&-3.7&3.1& 1.6 \cr
\hbox{Flows (bls \$)} \hfill &28&17&10&16&20&16&26&14&25&20&18&15 \cr
 \hbox{\bf 1998}  & & & & & & & & & & & & \cr
\hbox{NY+NAS (\%)} \hfill & &7.2&4.8&1.1&-2.6&3.2&-2.0& -16
&6.7&7.1&6.2&6.1 \cr
\hbox{Flows (bls \$)} \hfill &14&24&22&26&18&19&19&-11&6.2&2.4&13&3.2 \cr
  \hbox{\bf 1999}  & & & & & & & & & & & & \cr
\hbox{NY+NAS (\%)} \hfill & &-3.8&4.0&4.7&-2.1&5.3&-3.0&-0.65
&-2.3&6.2&4.3&9.3 \cr
\hbox{Flows (bls \$ )}\hfill &17&0.76&12&26&15&19&12&9.4&11&21&18&25 \cr
  \hbox{\bf 2000}  & & & & & & & & & & & & \cr
\hbox{NYSE  (\%)} \hfill & & -4.6&9.3&-0.46&-0.15&-0.31&-0.15& 5.3
&-1.6&0.45& -5.5 & \cr
\hbox{NASDAQ (\%)} \hfill & & 19.2& -2.6& -15.5& -11.9&16.6 &-5.0 & 11.7
&-12.7& -8.2&-22.9& \cr
\hbox{NY+NAS (\%)} \hfill & &4.5&5.0&-5.3&-3.6&5.2&-1.7& 7.4
& -5.0&-2.0&-9.8& \cr
\hbox{Flows (bls \$)} \hfill &41&53&39&34&17&22&17&23&17&19&5.7& \cr
\hbox{Bond 3-mo.(\%)} \hfill & 5.5&5.7&5.9&5.8&6.0&5.9&6.1& 6.3
&6.2& & & \cr
\hbox{Bond 10-yr.(\%)} \hfill & 6.7&6.5&6.3&6.0&6.4&6.1&6.0& 5.8
&5.8& & & \cr
} $$

\hrule
\vskip 0.5mm
\hrule

\vskip 2mm
Sources: Stock indexes: http://finance.yahoo.com; bond yields: 
http://www.fms.treas.gov/bulletin; flows: Mutual Fund Factbook (2000)

\normalsize

\end{table}

%%%%%%%%%%%%%%%%%%%%%%%%%%%%%%%

 % TABLE 1b

\begin{table}[htb]
\small

\centerline{\bf Table 1b \ Regression of monthly flows ($ f $ ) with respect
to price changes ($ \Delta p/p $): $ f =a\Delta p/p+b $}
\vskip 3mm
\hrule
\vskip 0.5mm
\hrule
\vskip 2mm
$$ \matrix{
\tvi  \hbox{Year}\hfill &a&b& \hbox{Correlation}&\hbox{Proportion} \alpha \cr
 & & & & \hbox{of speculative} \cr
 \tvi & & & & \hbox{investors} \cr
\noalign{\hrule}
\tvi 
1997 \hfill & 0.70 \pm 0.50 & 16 \pm 2& 0.67& 4.1 \% \cr
1998 \hfill & 0.70 \pm 0.97 & 11 \pm 6& 0.42& 5.7 \% \cr
1999 \hfill & 1.36 \pm 0.61 & 12 \pm 3& 0.82& 9.9 \% \cr
2000 \hfill & 1.36 \pm 1.35 & 25 \pm 7& 0.57& 5.1 \% \cr
} $$

\hrule
\vskip 0.5mm
\hrule

\normalsize

\end{table}

\qI{Evidence}

\qA{Short-term response (weekly fluctuations)}
\qun{Crash of August 1998}
Between 1 July and 30 September 1998 the Dow Jones index lost
13 percent (table 2). How did mf shareholders react?
As can be seen from table 1a the total inflow for July-September was
markedly lower than in the same quarter of 1997, but it was still
positive. There was only an outflow of money in one month, namely
11 billions of dollars in August, a reaction which is probably
connected with the fact that the most spectacular crash occurred
by the end of August: between 20 and 31 August the Dow Jones index
lost 12 percent. However this flow ($ f $ ) of 11 billions represented
only $ 11/2368 = 0.46 $ percent of the total assets of 
equity funds ($ A $ ).

 % TABLE 2

\begin{table}[htb]
\small

\centerline{\bf Table 2 \ Net outflows of money ($ f $) caused by stock
market crashes:} 
\centerline{\bf short term perspective, in percent of funds assets ($ A $ )}
\vskip 3mm
\hrule
\vskip 0.5mm
\hrule
\vskip 2mm

$$ \matrix{
\tvi \hbox{Year} & \hbox{Dates of crash} \hfill & \Delta p/p & f/A \cr
\tvi    &                 & \hbox{percent} & \hbox{percent} \cr
\noalign{\hrule}
\tvi
 1987 & \hbox{16 Oct. - 20 Oct.} \hfill & -23 & -2.1 \cr
 1990 & \hbox{15 Jul. - 15 Oct.} \hfill & -21 & -1.8 \cr
 1994 & \hbox{22 March - 6 April} \hfill & -7.0 & -0.28 \cr
 1994 & \hbox{Latin American stocks} \hfill & & \cr
 &   \hbox{15 Dec. - 31 Dec.} \hfill & -15 & -1.8 \cr
 1998 &\hbox{20 Aug. - 31 Aug.} \hfill & -12 & -0.46 \cr
} $$

\hrule
\vskip 0.5mm
\hrule

\vskip 2mm
Notes: The crash of 15-31 December concerned only funds comprising
Latin American stocks; it was due to the devaluation of the
Mexican peso on 20 Dec. 1994. The linear regression for the
five points in the table is: $ f/A = a\Delta p/p + b,
\quad a=0.11 \pm 0.06,\quad b=0.54 \pm 0.36 $ (the correlation
is 0.90) which gives $ \alpha = 17 $\ percent. 
\qL
Sources: Mutual Fund Factbook (2000), Marcis et al. (1995)., Reid (2000).

\normalsize

\end{table}

\qun{Crash of October 1987}
During the month of October 1987 the Dow Jones 
index lost 23 percent (table 2); in that month 
the reaction of mf shareholders resulted in an
outflow of cash representing 3.1 percent of equity funds assets. 
In November the outflow was only 0.5 percent and for December the
figure was almost the same. It is interesting to observe
that 70 percent of the October outflow occurred in the three
crash-days of October 16, 19, and 20 (Rea et al. 1996). 
Moreover a survey
conducted in November 1987 by the Investment Company Institute
shows that only 5 percent of the shareholders had sold shares
during the crash. A survey conducted after the more limited crash
of March 1994 similarly showed that only a small number (of the
order of 5 percent) of mf shareholders had sold shares.
\qL
In short, the picture which emerges 
from the above observations shows that a majority of 
shareholders are long-term investors who do not care
about short-term price fluctuations even when they assume
crash proportions. This is in agreement with the opinion expressed by
mutual fund managers: ``A vast majority of shareholders are seasoned
investors who do not intend to redeem [i.e. sell] shares in
response to adverse market developments'' (Rea et al. 1996).
\qL
Nevertheless, it would obviously be unreasonable
to imagine that mf shareholders will hold their shares 
no matter what happens in the market whatsoever.
To get a clearer view of that point we know analyze their
response to fluctuations extending over one or several years.

\qA{Long-term response (yearly fluctuations)}
So far we have focused on daily fluctuations; in the present paragraph
we consider fluctuations which have a typical duration of one or
two years. 
Yearly data for the decade 1990-2000 are summarized in table 3a;
they are not very conclusive however for there is almost no
correlation between price changes and flows whether expressed in
absolute or relative terms. A better procedure consists in slicing
the interval 1942-1990 into sub-intervals 
(of an average duration of about two years) during which there was a 
steady price increase (or decrease) on an annual basis. 
For each sub-interval we want to know whether shareholders have sold
or bought. The results are summarized in Fig.1 and table 3b. 
``Ideally'' one would expect a relationship of the form:
$$ f/A = a \Delta p /p + b $$

 % TABLE 3a

\begin{table}[htb]
\small

\centerline{\bf Table 3a \ Net new money invested in equity mutual funds}
\centerline{\bf versus stock price changes, annual data}

\vskip 3mm
\hrule
\vskip 0.5mm
\hrule
\vskip 1mm
$$ \matrix{
\tvi 
 & 1990&1991&1992&1993&1994&1995&1996&1997&1998&1999&2000 \cr
\noalign{\hrule}
\tvi
\hbox{NYSE Composite} \hfill &-7.7&22.5&4.7&7.9&-3.1&31.6&15.3&30.4
&16.8&8.3&0.9 \cr
\hbox{Flows, equity, (bl.\$ )} \hfill &13&38&74&115&102&107&182&186
&127&148&221 \cr
 & & & & & & & & & & & \cr
\hbox{Treas., 30-yr, yield (\% )} \hfill &8.5&7.8&7.4&6.5&7.1&6.8&6.5&6.4&5.5
&5.8&6.0 \cr
\hbox{Flows, bond, (bl.\$ )} \hfill &7&57&66&65&-55&-5&2.5&23&60&4.7
& \cr
} $$

\hrule
\vskip 0.5mm
\hrule

\vskip 2mm
Notes: The flows are deflated and expressed in 1990 dollars. The 
regression between price change and flows reads (correlation is 0.22):
$ f=a(\Delta p/p)+ b \quad a=1.1\pm 3.1 \quad b=106\pm 38 $ which 
leads to $ \alpha =1 $ percent. 
\qL
Sources: Price index: http://minneapolis.org/economy; NYSE Composite:
http://finance.yahoo.com; flows: Reid: Reid (2000); bond yield:
http://www.fms.treas.gov/bulletin.

\normalsize

\end{table}

%%%%%%%%%%%%%%%%%%%%%

 % TABLE 3b

\begin{table}[htb]
\small

\centerline{\bf Table 3b \ Relationship between changes in stock prices
($ \Delta p/p $)} 
\centerline{\bf and the net in- or out-flow of money ($ f/A $):
$ f/A = a\Delta p/p + b $, 1942-1990}

\vskip 3mm
\hrule
\vskip 0.5mm
\hrule
\vskip 2mm

$$ \matrix{
\tvi
\hbox{Observations} \hfill & \hbox{Number} & a & b & \hbox{Coef. of} & \cr
\hbox{} \hfill & \hbox{of} & & & \hbox{correl.}& \cr
\hbox{} \hfill & \hbox{observat.} & & & & \cr
\hbox{} \hfill &  & & &  &\cr
\noalign{\hrule}
\tvi
\hbox{All data} \hfill & 28 & 0.32 \pm 0.15 & 5.4 \pm 7.3 & \phantom{-}0.63 
&(0.34,0.81) \cr
\hbox{} \hfill &  & & &  &\cr
\hbox{\bf Bullish climate} \hfill &  & & & & \cr
\hbox{Price increases} \hfill & 10 & 0.60 \pm 0.29 & -10 \pm 10 
& \phantom{-}0.82 & (0.40,0.96) \cr
\hbox{Price decreases} \hfill & 10 & -0.14 \pm 1 & 5 \pm 7 
& -0.14 &(0.69,0.56) \cr
\hbox{} \hfill &  & & & & \cr
\hbox{\bf Bearish climate} \hfill &  & & &  &\cr
\hbox{Price increases} \hfill & 4 & -0.72 \pm 0.79 & 10 \pm 9 
& -0.78 & (-0.99, 0.72) \cr
\hbox{Price decreases} \hfill & 4 & 0 \pm 0.5 & -5 \pm 6 
& \phantom{-}0.02 & (-0.96, 0.96) \cr
} $$

\hrule
\vskip 0.5mm
\hrule

\vskip 2mm
Notes: The figures within parenthesis give the confidence interval
for the correlation at probability level 0.95. 
The regression for all data leads to $ \alpha =5.6 $ percent.
\qL

\normalsize

\end{table}

where $ a $ is positive and has (more or less) the same value
whether  $ \Delta p /p $ is positive or negative. Such a relationship
would mean that mf shareholders buy when prices increase and sell
when they fall. However Fig.1 shows that the regression line is not
the same for positive and negative price changes. For price 
increases there is an unambiguous positive correlation and regression,
whereas for price falls the regression coefficient 
is almost equal to zero. In other words, mf shareholders responded
to price increases by buying new shares, but to price falls
they reacted not by selling but by stopping to buy. 
\qL
One can go a step further. Consider the points in the 
$ \Delta p /p \geq 0 $ region; there is an obvious cluster of
outliers below the $ f/A = 0 $ axis; these points 
correspond to cases for which shareholders responded to a price
increase by selling, a rather puzzling behavior. A clue is the
observation that all these events occurred during the bear
market of 1970-1980. In other words in an overall bear climate
shareholders sell whether the annual price changes are positive or
negative. 
\qL
Is there a mirror-effect during bull markets, 
that is to say, do shareholders
buy during bull markets even when prices fall? We have already seen
that this was the case in 2000. In Fig.1 the points corresponding
to such a behavior would be in the quadrant: $ (\Delta p /p < 0,\
f/A > 0) $; not surprisingly all these cases occurred
during bull market periods (marked by up-going arrows). 
\qL
The regression coefficients corresponding to the anomalous
cases of 
quadrants 2 and 3 are defined with large error margins due to the small
number of cases in each sub-category; but, at least qualitatively 
there is no doubt as to the existence of the effects described
above. In a sense one is here is the same situation as in
neutrino physics in the late 1960s when the number of collision-events
for an experiment campaign lasting several weeks was of the order
of a few dozens. As in neutrino physics, in order to get better
accuracy one needs more events.

\qI{Conclusion}
According to economic theory the equilibrium price of a good
is determined by the intersection of the supply $ (S) $ and
demand $ (D) $ curves. For stock markets these curves are largely
unknown however.
On stock exchanges equilibrium prices are determined
by computer codes
and, for instance on the NYSE in the case of large transactions, by the 
intervention of the specialist in charge of that specific stock.
\qL
Note that 
the supply and demand curves cannot be reconstructed from 
the knowledge of 
the trading volume $ V = S+D $ and equilibrium prices. 
For that purpose one would need to know 
the bid-ask prices 
and bid-ask volumes, that is to say the prices and volumes
on the demand and supply sides (
people wanting to buy versus
people wanting to sell). Unfortunately these data are not
made public and thus
it is impossible to construct supply and demand curves
for a given stock. 
\qL
This is regrettable because models and simulations 
require specific assumptions to be made about the relationship between
excess-supply (i.e. $ s=S-D $) and the magnitude of the price changes
(sign and amplitude). For instance in the Cont-Bouchaud model
(Cont et al. 2000, Stauffer et al. 1999) 
it is assumed that the logarithm of the
price changes in proportion to excess-supply. Since that relationship
cannot be derived from observation it has to be conjectured 
which adds additional free parameters to the model. Note
however that, in contrast to other models, the Cont-Bouchaud
model does not make any specific assumptions regarding the
response of agents to price changes: the decisions to buy or to sell
are taken randomly. 
\qpar

In the present paper we have investigated the actual behavior of mf 
shareholders. 
In particular, we have seen that there 
is no simple relationship between price 
variation and propensity to buy or sell. 
What should be the next step? Obviously it would not be reasonable
to jump too quickly to the conclusion that {\it all} shareholders
have the same behavior as mf shareholders: inductive reasoning
based on only one case is fairly hazardous.
As we already 
emphasized, in order to make further progress one needs additional
data. It is not impossible that evidence for
other classes of shareholders may be available 
for instance in the publications
of insurance companies or commercial banks. Once one knows the
behavior of three or four classes of investors one will be
in a much better position to  make a reasonable inference about
the form of the excess-supply function $ s=s(p) $.
\qpar

Acknowledgment: The paper greatly benefited from 
Dietrich Stauffer's critical reading and advice; many thanks
to him. 

%%%%%%%%%%%%%%%%%%%%%%%%%%%%%%%%%%%%%%%%%%%

\null

\vfill \eject

\centerline{\bf \Large References}

\vskip 1cm

\qparr
Cont (R.), Bouchaud (J.-P.) 2000: Herd behavior and aggregate
fluctuations in financial markets. Macroeconomic Dynamics 4, 170.

\qparr
Marcis (R.), West (S.), Leonard-Chambers (V.) 1995: Mutual 
fund shareholders response to market disruptions.
Perspective 1,1 (July), published
by the Investment Company Institute.

\qparr
Rea (J.), Marcis (R.) 1996: Mutual fund shareholder activity
during U.S. stock market cycles 1944-1995. 
Perspective 2,2 (March), published
by the Investment Company Institute. 

\qparr
Reid (B.) 2000: The 1990s: A decade of expansion and change
in U.S. mutual fund industry. Perspective 6,3 (July), published
by the Investment Company Institute.

\qparr
Roehner (B.M.) 2001: Hidden collective factors in speculative
trading. Springer-Verlag. Berlin.

\qparr
Stauffer (D.), Sornette (D.) 1999: Self-organized percolation
model for stock market fluctuations.
Physica A 271, N3-4, 496-506.

\vskip 20mm

Figure caption
\qL
{\bf Fig.1 Net new money invested in mutual equity funds as a function
of stock price changes.} 
The dotted line
shows the regression line for all points together; it can
be seen that there are two clusters of outliers. The arrows 
indicate the bullish or bearish state of the market, as defined
in the inset which shows the evolution of the Standard and Poor's
index (a moving average was performed in order to highlight the 
trends). 
That additional variable accounts for the anomalous points
in quadrants 2 ($ x<0, y>0 $) and 3 ($ x>0, y<0 $). 
Source: Rea et al. (1996). 

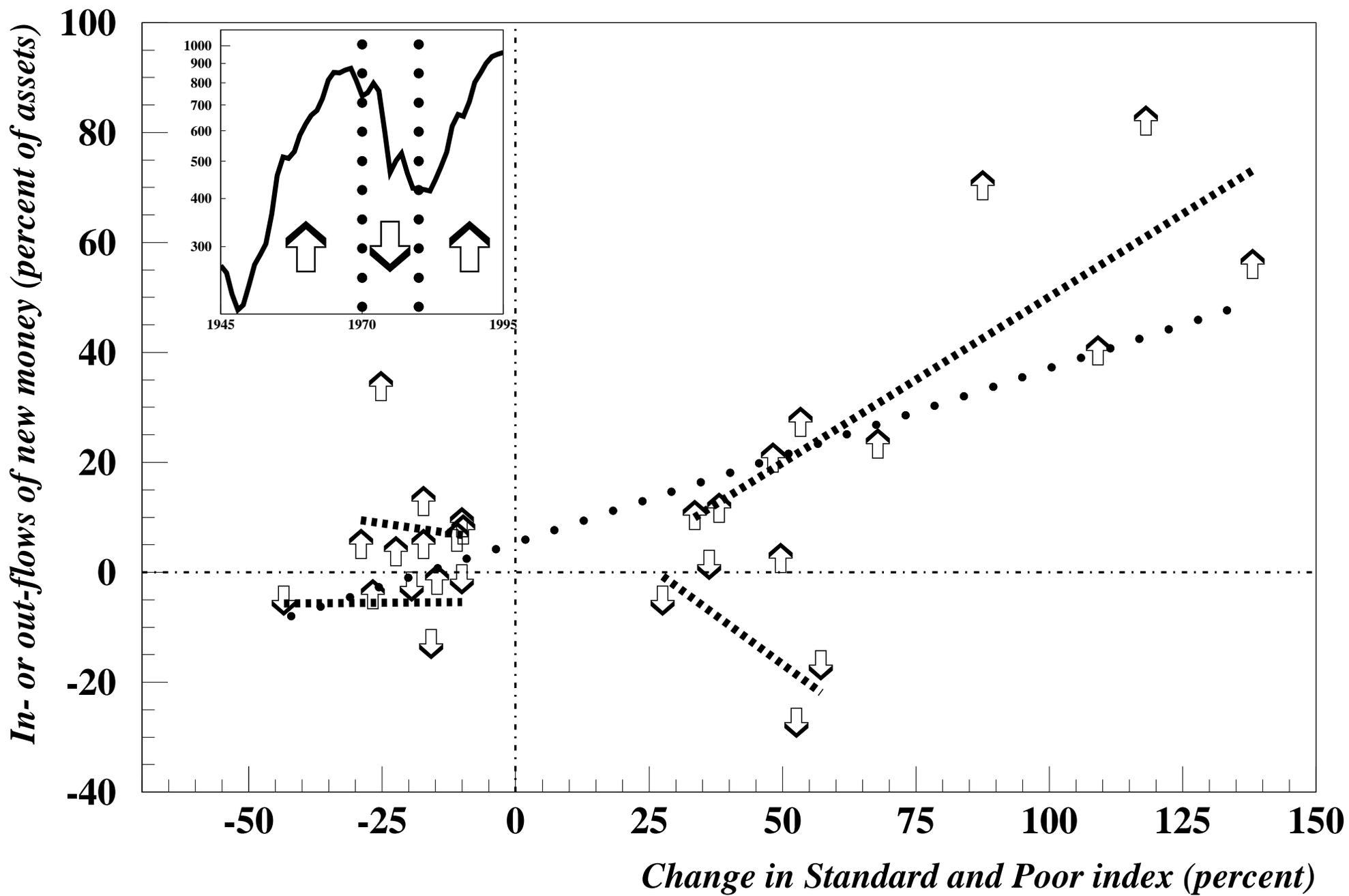
\end{document}